\documentclass[sigconf,dvipsnames]{acmart} 

\usepackage{booktabs} \usepackage{xcolor}
\usepackage{siunitx}
\usepackage{multirow}
\usepackage{balance}
\usepackage[font=rm]{caption}
\usepackage{algorithm}
\usepackage[noend]{algpseudocode}
\usepackage{verbatim}
\usepackage{graphicx}
\usepackage{tabularx}
\usepackage{enumitem}
\usepackage{shortvrb}

\setlist[itemize,1]{leftmargin=3mm,itemsep=1mm}
\setlist[enumerate,1]{leftmargin=3mm,itemsep=1mm}

\newcommand{\mb}[1]{{\mbox{$#1$~MiB}}}
\newcommand{\gb}[1]{{\mbox{$#1$~GiB}}}

\newcommand{\bol}[1]{\mbox{\bf{#1}}}
\newcommand{\var}[1]{\mbox{\emph{#1}}}

\newcommand{\myparagraph}[1]{\paragraph*{\hspace*{-\parindent}\normalsize\bf#1}}
\newcommand{\mycaption}[1]{\caption{{\rm{#1}}}}

\usepackage{xcolor}
\definecolor{tealgreen}{rgb}{0.0, 0.51, 0.5}

\newcommand{\figheight}{50mm}

\copyrightyear{2026}
\acmYear{2026}
\setcopyright{cc}
\setcctype{by}
\acmConference[]{This work is, as of April 2026}{}{unpublished.}
\acmBooktitle{}
\acmISBN{}
\acmDOI{}

\makeatletter \gdef\@copyrightpermission{
\begin{minipage}{0.3\columnwidth}\href{https://creativecommons.org/licenses/by/4.0/}{\includegraphics[width=0.90\textwidth]{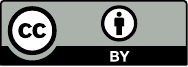}}\end{minipage}\hfill \begin{minipage}{0.7\columnwidth}
\href{https://creativecommons.org/licenses/by/4.0/}{This work is licensed under a Creative Commons Attribution
International 4.0 License.}\end{minipage}\vspace{5pt}}

\begin{document}
\fancyhead{}
\title{Speed Thrills:
Visceral Demonstrations That\\
Get Students Excited About Efficient Algorithms}

\author{Alistair Moffat}
\orcid{0000-0002-6638-0232}
\affiliation{\institution{The University of Melbourne}
  \city{Melbourne} 
  \country{Australia} 
}
\email{ammoffat@unimelb.edu.au}

\author{David Hawking}
\orcid{0000-0002-3704-5398}
\affiliation{\institution{Australian National University}
  \city{Canberra} 
  \country{Australia} 
}
\email{david.hawking@acm.org}


\begin{abstract}
We address the problem of motivating students in Data Structures and
Algorithms courses by presenting two simple problems that each have a
series of improvements to a basic algorithm, leading to spectacular
decreases in runtimes.
Coining a new term, we refer to such sequences as being
``{\emph{thrills of algorithms}}''.
Seeing runtimes drop from an estimate of days (or even years) to just
a few seconds has a visceral impact which conveys the importance of
efficient algorithms in a way unlikely to be forgotten.
The demonstrations are particularly compelling because they can be
performed live in class on the lecturer's laptop.
To assist staff teaching such courses we provide detailed pseudocode
descriptions and complexity analyses for the various methods, and can
supply implementations on request.
\end{abstract}
 
\maketitle

\section{Introduction}

One of the most important lessons a budding computer scientist must
acquire is the knowledge of algorithmic problem solving, and an
understanding that a ``correct'' solution to a computational problem
may not necessarily be a ``useful'' one -- that a program that
requires aeons to declare a solution is not any better than not
having a program at all.
This lesson is often associated with courses in data structures and
algorithms, and presented via mathematical definitions of ``big Oh''
and similar notations, topics which students can find dry and
challenging.
Our experience through several decades of such subjects is that this
lesson benefits enormously from being made visceral, and experienced
via the senses as well as by logic.

To this end we routinely make use of two simple computational
problems that have a number of attractive properties: both have
simple descriptions that even novice programmers are able to grasp
and then consider solutions to; both have obvious solutions that are
viable when tested at small scales, but with execution times that
grow rapidly when the problem size is increased; and both have a
range of solutions of increasing asymptotic efficiency that can be
understood with only a little more effort.
In writing this paper we pondered a suitable collective noun for a
such a hierarchy of algorithms all addressing the same problem, and
eventually settled on ``thrill'', as in, ``{\emph{a thrill of
algorithms}}''.

In particular, we argue that being able to present a sequence of
approaches provides the visceral understanding that is so important.
In the ``normal'' world, a breakthrough that saves $50$\% of some
resource would be a remarkable improvement.
For example, think of the recognition that would be showered on an
aeronautical engineer if they were able to save $50$\% of the fuel
cost of intercontinental flights.
Indeed, even a $5$\% savings would bring substantial professional
recognition.

Now consider a computational task of size $n$ and a straight-forward
approach to solving it.
Imagine executing that program on problem instances of increasing
size with you delivering some pre-planned commentary while your
students watch: ``let's try $n=1000$, wow, that was fast; let's try
$n=2000$, still fast; $n=4000$, was that a very small `blink'
occurred before the answer came, maybe we can try and measure the
time taken.''
Then imagine enabling some kind of automatic timing, and starting a
spreadsheet to construct a record of the running times as $n$ is
increased.
And segueing from measurement alone, into predictions while you do~so.

Now suppose that the $n=10{,}000$ instance is measured as requiring
$0.03$~seconds; and that when you get to it the $n=100{,}000$
instance takes around $3$~seconds, which $100$~times longer.
Now you initiate the $n=1{,}000{,}000$ execution, and then continue
with the planned script while it is running, using the spreadsheet to
analyze the running times observed so far, explaining as you go that
once you have a model the accounts for the observations to date,
you'll be able to start making predictions with it.
That ``sales patter'' needs to take around $5$ minutes, and needs to
conclude that the $n=1{,}000{,}000$ run will take something like
$300$ seconds (which is the $5$ minutes you have been talking for, of
course).
If you get the timing right the $n=1{,}000{,}000$ run will finish
right on cue when you need it, and take $100$ times longer than the
$n=100{,}000$ one.

\begin{figure*}
\centering
\includegraphics[width=0.90\textwidth]{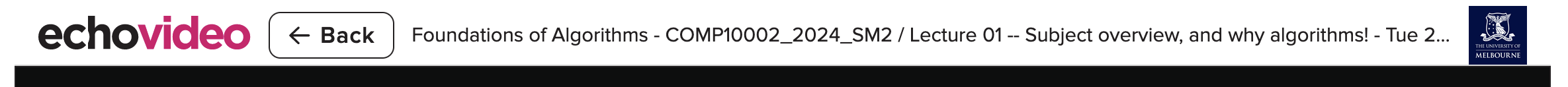}
\includegraphics[width=0.90\textwidth]{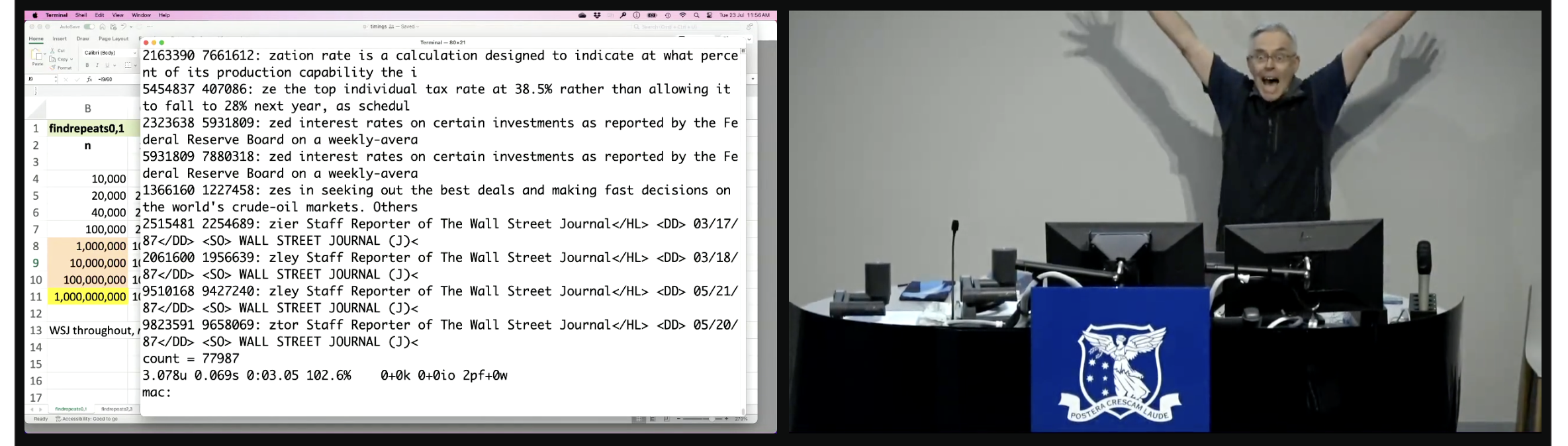}
\mycaption{One of the authors reaches the visceral ``tadaa'' moment
in a lecture when a ``flight from Melbourne to Shanghai'' (Repeats
Method~1, $n=10^7$, $m=100$, extrapolated to take nine hours based on
an execution of $n=10^6$ needing $320$ seconds, see
Section~\ref{sec-repeats}) is accomplished in just three seconds
using a better algorithm (Repeats Method~2).
\label{fig-tadaa}}
\end{figure*}

Then initiate (why stop at $1{,}000{,}000$!)\ the $n=10{,}000{,}000$
execution in the terminal window, and return to the spreadsheet.
Based on the observation already accumulated, it should now seem
plausible to expect a problem of that size to take $100$ times $5$
minutes, with that estimate allowing you an exclamation of ``wow,
gee, that's $9$ hours, that's how long it takes to fly from Melbourne
to Shanghai''.
And you then explain that you need to control-C the execution,
because no-one in the room wants the lecture to last another $9$
hours.
And then, still in the spreadsheet, point out that invoking another
factor of ten on $n$ leads to another (predicted) factor of $100$ on
the execution time, at which point you can pretend to pull your hair
out and cry ``gee, wow, that would be more than a {\emph{month}} of
computation to handle $n=100{,}000{,}000$, I guess no point starting
that one, none of us want to wait that long''.
Followed by ``and, guess what, if $n$ is a billion it's going to take
10 years.
Some of you might still be at the University then, but I probably
won't be!''

Then imagine switching to a different mechanism with ``here's a
different program I prepared earlier, I'm not going to tell you yet
how it works, you just need to trust me'', and asking the class, ``do
you think this new program will be able to solve an
$n=10{,}000{,}000$ instance in less than an hour?
``Could it be $9$ times faster than the one we just saw?'', you say.
``That would be like flying from Melbourne to Shanghai in just one
hour, wouldn't that be amazing!''
Then keep right on going: ``what about less than $10$ minutes?
Do you think that might be possible?
Less than $1$ minute??
Will you give me a standing ovation if I can do $n=10{,}000{,}000$ in
less than a minute???''

Well, with the right choice of ``before'' and ``after'' algorithms,
you'll get your standing ovation for sure, because the difference
between an $O(n^2)$ algorithm (which is what the first one is,
obviously) and an $O(n\log n)$ algorithm is so compellingly
substantial that the latter will -- even allowing for a likely higher
constant coefficient -- result in an execution time that is not just
certainly less than a minute for $n=10{,}000{,}000$, but quite
probably {\emph{less than ten seconds}}.
Finally, you tell the students: ``by the end of this subject you'll
know how I did that, you'd better keep coming to class!''
That's what we mean by {\emph{visceral}} -- something that is felt
not just intellectually by the brain, but also as a metaphorical
``punch in the stomach''.

What problems are suited to for this type of demonstration?
We describe two favorites that we have been using for many years in
our classes.
The first problem is that of counting primes, answering the question
``how many primes are there less than $n$?''\ when presented with an
integer $n$.
The second question is a string matching one: ``given a string of $n$
total characters, how many substrings of length $m$ recur within
it?''.
Figure~\ref{fig-tadaa} shows a lecture recording screenshot of the
critical ``tadaa'' moment in a recent class, when a nine hour
``flight from Melbourne to Shanghai'' was reduced to to just $3$
seconds, right on cue at the end of the first lecture in the subject.
The $450$ students that were present took the hint that had been
seeded into the build-up, and rose in a standing ovation.

Prime counting and repeat counting both support a {\emph{thrill of
algorithms}}, a sequence of approaches in which each improves on its
predecessor; and with each increment in the sequence simple enough
that it can be readily understood by first-year programming students.
These are the students that we wish to convey the over-riding
``algorithms are fun'' mantra to, and touch the minds of.

The next two sections describe these two problems in more detail, and
the ``thrill of algorithms'' that that can be presented to students.
We also describe other parts of the ``about algorithms and their
differences'' lesson that is so important for computing students to
understand.
A repository of C code is available if you are interested in adopting
this ``live visceral approach'' in your own algorithms classes.
 \section{Counting Primes}
\label{sec-primes}

The problem considered in this section is the well-known one of
counting primes.
None of these solutions is novel in any way; indeed, the internet is
full of programs that solve this task, and while prime counting makes
for a compelling lecture demonstration, it almost certainly should
{\emph{not}} be used as a class assignment.

\myparagraph{Primes Method 0}

The ``starting point'' mechanism for prime counting, dubbed
``Method~0'', is embarrassingly inefficient, and we do not present
pseudocode.
In this approach every factor $2\dots c-1$ for each candidate prime
$2\le c < n$ is tried, regardless of the outcome of previous
comparisons.
Checking a candidate $c$ then requires $c-2$ divisions, and hence
testing all values $1\le c \le n$ requires $(n-1)(n-2)/2$ divisions.
Overall running time for counting the primes up to $n$ thus grows
quadratically: if $n$ doubles, the execution time goes up by a factor
of four; and if $n$ increases by a factor of ten, the running time
increases by a factor of~$100$.

\myparagraph{Primes Method 1}

\begin{algorithm}[t]
\begin{algorithmic}[1]
\State set $\var{nprimes} \leftarrow 0$
\State // check all candidates in range
\For{$c \leftarrow 2$ {\bol{to}} $n-1$}
  \State set $\var{isprime} \leftarrow 1$
  \State // check every possible divisor
  \For{$d \leftarrow 2$ {\bol{to}} $c-1$}
    \If{$c ~\bol{mod}~ d = 0$}
      \State set $\var{isprime} \leftarrow 0$
      \State // end inner loop if a factor is found
      \State \bol{break}
        \label{stp-primes1-break}
    \EndIf
  \EndFor
  \State set $\var{nprimes} \leftarrow \var{nprimes}+\var{isprime}$
\EndFor
\State \Return \var{nprimes}
\end{algorithmic}
 \caption{Counting the primes less than $n$, Method~1.
\label{alg-primes1}}
\end{algorithm}

Most students will quickly grasp that if a factor is found then there
is no need to continue dividing, and that identification of any one
factor for some candidate $c$ is enough to demonstrate non-primality.
This observation leads to Method~1, described in
Algorithm~\ref{alg-primes1}, in which a {\bol{break}} statement
allows early exit from the inner loop as soon as a factor is found,
see step~\ref{stp-primes1-break} (that is, if that statement is
removed, we have Method~0).

The {\bol{break}} statement means any candidate $c$ that is a
multiple of two is discarded after one division, any odd numbers that
are multiples of three are discarded after two divisions, and so on.
This simple alteration to the code creates a considerable speed
boost, and already provides an educational ``thrill''.
For example, on a MacBook Pro M3 (2024) laptop, Method~0 requires
$2.5$ seconds when $n=10^5$, whereas Method~1 requires $0.26$
seconds, almost ten times faster.

This first simple acceleration also allows discussion of programmer
responsibilities and ethics, covering both the need for code
correctness, and also the need to understand execution times so as to
be environmentally responsible and reduce the carbon footprint (via
server electricity and cooling costs in data centers) of
computational tasks.

As was noted, the growth in execution time of Method~0 is clearly
quadratic because of the nested loop structure.
But the early exit possibility in Method~1 raise the question: how
does the running time now grow as $n$ increases?
If a candidate $c$ is a prime, then $c-2$ divisions will still be
required.
But primes become (slowly) increasingly sparse as $n$ increases, and
the density of primes in the interval $[1\ldots n)$ is
approximately\footnote{See
{\url{https://en.wikipedia.org/wiki/Prime_number_theorem}}.}
$1/\ln n$.
At this point in the lecture delivery you might also note that if a
number is composite, then there must be a factor $d$ that satisfies
$d \le \sqrt{c}$.
That means that the cost of confirming all of the primes up to $n$
using Method~1 is $O(n^2/\log n)$ (same as with Method~0), and that
the Method~1 cost of eliminating the composite numbers is
$O(n^{1.5})$ which is less, even though the composites greatly
outnumber the primes by a factor of $\ln n$.
Method~1 thus has an overall time cost that is $O(n^2/\log n)$.

\begin{table}[t]
\centering
\sisetup{
group-separator = {,},
round-mode = places,
round-precision = 1,
table-format=6.1,
}\begin{tabular}{ccccc}
\toprule
$n$
	& Method 0
		& Method 1
			& Method 2
				& Method 3
\\
\midrule
$10^3$
	& 497,503
		& 78,021
			& 5,287
				& 2,800
\\[1ex]
$10^4$
	& ${}\times 100.5$
		& ${}\times 74.0$
			& ${}\times 22.2$
				& ${}\times 15.6$
\\
$10^5$
	& ${}\times 100.0$
		& ${}\times 78.8$
			& ${}\times 23.4$
				& ${}\times 17.0$
\\
$10^6$
	& ${}\times 100.0$
		& ${}\times 82.5$
			& ${}\times 24.7$
				& ${}\times 18.7$
\\
$10^7$
	& ${}\times 100.0$
		& ${}\times 85.3$
			& ${}\times 25.8$
				& ${}\times 20.5$
\\[1ex]
$=$
	& $5.0 \times 10^{13}$
		& $3.2 \times 10^{12}$
			& $1.7 \times 10^9$
				& $2.9 \times 10^8$
\\
\bottomrule
\end{tabular}
 \mycaption{Divisions required as $n$ increases for four different
prime counting methods.
The first data row shows the absolute count of {\bol{mod}} operations
needed when $n=10^{3}=1000$.
The next four rows show the multiplicative growth factor for the
operation counts as $n$ increase by four factors of ten,
to reach $n=10^7=10{,}000{,}000$.
The last row then shows raw counts of {\bol{mod}} operations again.
\label{tbl-primes-divisions}}
\end{table}

An empirical exploration that tabulates running times or operation
counts as $n$ grows adds tangible numbers that complement these
abstract relationships.
In particular, calculating ratios between successive measurements
reinforces the importance of asymptotic growth rates.
As an example of what can readily be built up live during a lecture,
the first two data columns in Table~\ref{tbl-primes-divisions} list
the number of {\bf{mod}} operations required by Methods~0 and~1
across a range of values of $n$, expressed in the first row as raw
counts, and then in four further rows as growth multipliers as $n$
increases by factors of $10$.
As expected, the growth ratios for Method~0 are consistently
``${}\times100$'', in line with the $O(n^2)$ expectation.
On the other hand the growth ratios for Method~1 are consistently
smaller than $100$, a result of the $\log n$ in the denominator,
which serves as a dampener on the previous quadratic asymptotic
growth rate.
That is, Method~1 is not just faster than Method~0 by a fixed factor,
but by a ratio that grows as $n$ increases, meaning that the
``thrill'' factor of running them in a side-by-side shootout will get
more and more impressive as $n$ increases.

\myparagraph{Primes Method 2}

\begin{algorithm}[t]
\begin{algorithmic}[1]
\State set $\var{nprimes} \leftarrow 0$
\State // check all candidates in range
\For{$c \leftarrow 2$ {\bol{to}} $n-1$}
  \State set $\var{isprime} \leftarrow 1$
  \State // check possible divisors up to the square root
  \For{$d \leftarrow 2$ {\bol{to}} $c-1$ {\bol{while}} $d\times d \le c$}
      \label{stp-primes2-sqrt}
    \If{$c ~\bol{mod}~ d = 0$}
      \State set $\var{isprime} \leftarrow 0$
      \State // end inner loop if a factor is found
      \State \bol{break}
    \EndIf
  \EndFor
  \State set $\var{nprimes} \leftarrow \var{nprimes}+\var{isprime}$
\EndFor
\State \Return \var{nprimes}
\end{algorithmic}
 \caption{Counting the primes less than $n$, Method~2.
\label{alg-primes2}}
\end{algorithm}

Your students may well be champing at the bit by now, wondering why
you are talking about saving a factor of $\log n$, when there is a
much larger saving available to be harvested, shown in
Algorithm~\ref{alg-primes2}.
This version is a simple consequence of the analysis of Method~1 --
if a composite number $c$ must have a factor that is less than
$\sqrt{c}$, then, conversely, $c$ can be declared prime if no such
factor is found.
And with the checking of primes being the dominant element in the
running time of Method~1, this observation allows further reductions
in execution cost.
In Method~2 the inner loop over candidates $c$ is thus ended as soon
as $d^2>c$ (step~\ref{stp-primes2-sqrt}), meaning that when a number
$c$ {\emph{is}} a prime, only $\sqrt{c}$ divisions are required to
establish that fact, rather than the previous $c-2$.

That projected saving in running time is readily demonstrated.
Method~1 takes $18.6$ seconds to count the primes up to $n=10^6$ and
a trudging $26$ minutes for $n=10^7$; Method~2 counts the primes less
than $n=10^6$ in just $0.09$ seconds, and the primes less than
$n=10^7$ in a piffling $1.23$ seconds, results that are, as required,
wonderfully visceral.
Table~\ref{tbl-primes-divisions} further quantifies the dramatic
gains that are achieved.

It is also interesting to consider the asymptotic performance of
Method~2.
With approximately $n/\ln n$ primes to be verified, and $O(\sqrt{n})$
time required for each, the net cost of checking the prime numbers up
to $n$ is now $O(n^{1.5}/\log n)$.
But what about the process of eliminating the composites, making up
the majority of the values that get tested?
The discussion of Method~1 argued that this cost was $O(n^{1.5})$,
which in Method~1 was not dominant and could be ignored.
But that was an easy-to-motivate loose bound, used just to show that
in Method~1 the prime verification cost was dominant.
Method~2 greatly reduces the verification cost, meaning that
composite elimination is no longer a secondary component that can be
absorbed.
In fact, in Method~2 the cost of the divisions that identify the
composite numbers is also $O(n^{1.5}/\log n)$, a claim that will be
demonstrated after we have considered Method~3.
Table~\ref{tbl-prime-asymptotics} summarizes the running costs of the
three algorithms discussed to this point, separating the cost of
``verifying primes'' from the cost of ``eliminating composites''.

\begin{table}[t]
\centering
\begin{tabular}{l ccc}
\toprule
\multirow{2}{*}{Method}
	& Prime 
		& Composite 
			& Overall
\\
	& verification
		& elimination
			& analysis
\\
\midrule
Primes 0
	& $O(n^2/\log n)$
		& $O(n^2)$
			& $O(n^2)$
\\
Primes 1
	& $O(n^2/\log n)$
		& $O(n^{1.5})$
			& $O(n^2/\log n)$
\\
Primes 2
	& $O(n^{1.5}/\log n)$
		& $O(n^{1.5}/\log n)$
			& $O(n^{1.5}/\log n)$
\\
Primes 3
	& $O(n^{1.5}/\log^2 n)$
		& $O(n^{1.5}/\log^2 n)$
			& $O(n^{1.5}/\log^2 n)$
\\
Primes 4/5
	& $O(n)$
		& $O(n \log\log n)$
			& $O(n \log\log n)$
\\
\bottomrule
\end{tabular}
 \mycaption{Asymptotic cost of counting primes, split into two
components: the time taken to verify the set of approximately $n/\ln
n$ values less than $n$ that are prime numbers; and the cost of
demonstrating that each of the other $n(1-1/\ln n)$ numbers are
composite.
Note that while all bounds provided are correct, not all of them are
``exact'' relationships.
\label{tbl-prime-asymptotics}}
\end{table}

\myparagraph{Primes Method 3}

Your smart students are again likely to have their hands waving in
the air, franticly seeking to catch your eye.
They'll be wanting to tell you that check-dividing by (for example)
four or any other multiple of two is pointless, because no new
composite values will be identified.
Likewise, once three has been used as a potential divisor, there is
no point checking any multiple of three.
This idea then leads to Method~3, described in
Algorithm~\ref{alg-primes3}, which maintains an array of primes seen
so far, using them (and only them) as possible divisors $d$ against
each candidate $c$.

\begin{algorithm}[t]
\begin{algorithmic}[1]
\State // initialize an array of known primes
\State set $\var{nprimes} \leftarrow 0$ and create
	an array $\var{primes}[]$
\State // check all candidates in range
\For{$c \leftarrow 2$ {\bol{to}} $n-1$}
  \State set $\var{isprime} \leftarrow 1$
  \State // check all possible prime divisors \dots
  \For{$i \leftarrow 0$ {\bol{to}} $\var{nprimes}-1$}
    \State set $d \leftarrow \var{primes}[i]$
    \If{$d^2 > c$}
      \label{stp-primes3-sqrt}
      \State // \dots but stop when square root has been passed
      \State \bol{break}
    \EndIf
    \If{$c ~\bol{mod}~ d = 0$}
      \State set $\var{isprime} \leftarrow 0$
      \State // end inner loop if a factor is found
      \State \bol{break}
    \EndIf
  \EndFor
  \If{$\var{isprime} = 1$}
    \State // append new prime $c$ to the array of known primes
    \State set $\var{primes}[\var{nprimes}] = c$
      \label{alg-primes-store}
    \State set $\var{nprimes} \leftarrow \var{nprimes}+1$
  \EndIf
\EndFor
\State \Return \var{nprimes}
\end{algorithmic}
 \caption{Counting the primes less than $n$, Method~3.
An array of verified primes less than $n$ is created.
\label{alg-primes3}}
\end{algorithm}

The drawback of this approach is the need for the array
$\var{primes}[]$.
In a dynamic memory environment it can be declared small and then
resized on demand, and if doubled each time it becomes full, will
require at most twice as much space as the number of primes being
identified, which, as already noted, is approximately $n/\ln n$.
For example, if $n=10^{9}$ (one billion), around $5 \times 10^7$
integers would need to be stored, meaning that a resizable array of
not more than $\mb{800}$ will be required (assuming the use of
eight-byte integers).

Like Algorithm~\ref{alg-primes2}, Algorithm~\ref{alg-primes3} only
tests divisors up to $\sqrt{c}$ for each candidate $c$
(step~\ref{stp-primes3-sqrt}).
That is, if we have $n'=\lceil\sqrt{c}\rceil$ for some candidate $c$,
then the Prime Number Theorem asserts that a total of approximately
$n'/\ln n' =\sqrt{n}/\ln(\sqrt{n}) = 2\sqrt{n}/\ln n$ ``remainder
after division'' operations are needed to verify each of the $n/\ln
n$ prime numbers, saving another factor of $\log n$ off the running
time.
The Primes~3 ``prime verification'' entry in
Table~\ref{tbl-prime-asymptotics} reflects that cost.

The middle column in Table~\ref{tbl-prime-asymptotics} claims that
the cost of eliminating the composite numbers is also less than the
same bound.
While the details are somewhat complex and beyond what is possible to
explain in a ``this is why you need to study algorithms'' lecture,
the total number of
{\bol{mod}} operations across both categories can be shown to grow as
$O(n^{1.5}/\log^2 n)$, thereby covering all of the bounds shown in
the ``Primes~3'' row in Table~\ref{tbl-prime-asymptotics}.

\myparagraph{Primes Method 4}

\begin{algorithm}[t]
\begin{algorithmic}[1]
\State // initialize the sieve
\State create an array $\var{isprime}[2\ldots n-1]$
\For{$c \leftarrow 2$ {\bol{to}} $n-1$}
  \State set $\var{isprime}[c] \leftarrow 1$
\EndFor
\State set $\var{nprimes} \leftarrow 0$
\State // now step through the sieve, looking for survivors
\For{$v \leftarrow 2$ {\bol{to}} $n-1$}
  \label{stp-primes4-mainloop}
  \If{$\var{isprime}[v] = 1$}
    \label{stp-primes4-scanning}
    \State // $v$ has not been eliminated, so must be prime
    \State set $\var{nprimes} \leftarrow \var{nprimes}+1$
    \State // now eliminate all remaining multiples of $v$
    \State set $c \leftarrow v^2$
      \label{stp-primes4-cancelfrom}
    \While{$c < n$}
      \label{stp-primes4-eliminate-1}
      \State set $\var{isprime}[c] \leftarrow 0$
      \State set $c \leftarrow c + v$
        \label{stp-primes4-eliminate-2}
    \EndWhile
  \EndIf
\EndFor
\State \Return \var{nprimes}
\end{algorithmic}
 \caption{Counting the primes less than $n$, Method~4 (Sieve of
Eratosthenes).
An array of all possible candidates is reduced until only the primes
survive and remain uneliminated.
\label{alg-primes4}}
\end{algorithm}

An obvious question then arises: is there any remaining redundancy?
Or have we hit the wall in terms of reducing running times for this
problem?
Algorithm~\ref{alg-primes4} shows that considerable further gains are
possible.
It presents a famous approach attributed to the Greek mathematician
Eratosthenes\footnote{See
{\url{https://en.wikipedia.org/wiki/Sieve_of_Eratosthenes}}.}
in which the switches from checking individual candidates to checking
the entire range of candidates in a holistic manner.
That re-orientation allows another huge saving in execution time.

Method~4 employs a array containing one element per candidate, with
all entries initially set to ``true'' (one).
At each iteration of the main loop (step~\ref{stp-primes4-mainloop}),
if a ``still true'' array element is reached, then the corresponding
index $v$ must be prime, and all multiples of $v$ starting from $v^2$
are set to ``false'' in a leapfrogging elimination step.
The process then continues, stepping past already-eliminated values,
and pausing at each non-eliminated value, declaring it to be prime,
and eliminating its multiples.

This mechanism also requires an array, and in
Algorithm~\ref{alg-primes4} it contains $n$ values rather than the
$n/\ln n$ values required by Algorithm~\ref{alg-primes3}.
In Algorithm~\ref{alg-primes3} each array element is a prime number,
and at the scale we are considering in these implementations, that
means an eight-byte {\tt{uint64\_t}} variable.\footnote{But worth
noting is that Algorithm~\ref{alg-primes3} only access the array
$\var{primes}[]$ sequentially.
So each stored value could also be a {\emph{difference}} between
consecutive primes, in which case {\tt{uint16\_t}} values would
suffice.
The sequence of differences could also then be compressed using any
available integer compression technique {\citep[Chapter 3]{wmb99mg}}
to further reduce the space requirement, albeit at the expense of
extra execution time to decode values out of the compressed bit-array
as each candidate is being considered.}
On the other hand, in Algorithm~\ref{alg-primes4} the items being
stored are binary values and $\var{isprime}[]$ could be an array of
single bits.
Indeed, if all even numbers beyond two are automatically excluded and
are not stored, the array can be configured so that $16$ values in
the target range are represented in packed form in each array byte,
with $128$ candidates stored in each eight-byte {\tt{uint64\_t}}
variable.
Such packing makes the array smaller than the {\tt{uint64\_t}} array
assumed in Algorithm~\ref{alg-primes3} for all plausible values
of~$n$.
Despite these possibilities, the implementation of
Algorithm~\ref{alg-primes4} that accompanies this paper employs an
array of bytes rather than making use of a denser packing; why this
is a reasonable choice will emerge shortly, when Method~5 is
discussed.

One aspect of Algorithm~\ref{alg-primes4} that is not captured in the
pseudocode is the need to be careful with integer precision and
integer operations.
In particular, the computation at step~\ref{stp-primes4-cancelfrom}
will overflow even 64-bit arithmetic once $v$ exceeds the limits of
32-bit integers.

What about the running time?
Isn't the cost still the same as Method~3, just with a different
ordering of the elemental operations?
Surprisingly, the reordered operation sequence makes a huge
difference.
The ``prime verification'' column of
Table~\ref{tbl-prime-asymptotics} shows $O(n)$ time, and that
reflects the cost of the testing operations that take place at
step~\ref{stp-primes4-scanning} of Algorithm~\ref{alg-primes4}.
The ``candidate elimination'' cost then accounts for the loop from
step~\ref{stp-primes4-eliminate-1} to
step~\ref{stp-primes4-eliminate-2}.
That loop iterates $n/p_i$ times for each prime number $p_i$ that
gets identified, meaning that in total it iterates
\[
	\sum_{i=1}^{p_i<n} \frac{n}{p_i}
	= n \cdot \sum_{i=1}^{p_i<n} \frac{1}{p_i}
	= n (k + \ln \ln n)
\]
times, with the second summation being the prime harmonic sequence,
and $k$ a small constant\footnote{See
{\url{https://en.wikipedia.org/wiki/Divergence_of_the_sum_of_the_reciprocals_of_the_primes}}.}.
That is, Algorithm~\ref{alg-primes4} outperforms
Algorithm~\ref{alg-primes3} by another factor of
$O(n^{0.5}/((\log^2 n)(\log\log n)))$,
a ratio that once again translates into visceral improvement.
For example, where Method~3 takes $106$ seconds to compute the primes
up to $n=10^{9}$, Method~4 requires just $6.5$ seconds.
And Method 2 would have taken around $45$ {\emph{minutes}}.

With this additional gain, maybe even your smartest students will now
suspect that the end of the development has been reached, and that
this particular ``thrill of algorithms'' has reached its crescendo.

\myparagraph{Primes Method 5}

``But wait,'' you cry, ``there's more!'',\ setting your students'
senses tingling all over again.

While Method~4 is very efficient, just a tiny smidgen greater than
linear in $n$, it also requires rather a lot of memory space, an
amount linear in $n$ (which remains the case even if $16$ candidates
are packed into each byte).
That severely limits its usefulness.
For example, to compute the primes up to one trillion ($n=10^{12}$)
would require $10^{12}/16$ bytes, or $\gb{58}$, more than even
relatively well-configured laptops can offer.
And if a simple representation that uses an array of {\tt{uint8\_t}}
values was used, one byte per candidate, the situation would be
untenable.

\begin{algorithm}[t]
\begin{algorithmic}[1]
\State // create the set of primes needed to support the sieve
\State use Method~4 to calculate the $s\approx 2\sqrt{n}/\ln n$ primes
less than
\State \qquad or equal to $\sqrt{n}$, and store them in
	$\var{primes}[0\ldots s-1]$
\For{$i \leftarrow 0$ {\bol{to}} $s-1$}
  \State set $\var{next}[i] \leftarrow \var{primes}[i]\times\var{primes}[i]$
    \label{stp-primes5-setnext}
\EndFor
\State create an array $\var{isprime}[0\ldots\var{segsize}-1]$
\State set $\var{bot} \leftarrow 0$ and $\var{top} \leftarrow \var{segsize}$
\State // iterate over segments, each one from $\var{bot}$ to $\var{top}-1$
\While{$\var{bot} < n$}
  \State // reset the sieve array, ready for this segment
  \For{$c \leftarrow 0$ {\bol{to}} $\var{segsize}-1$}
    \State set $\var{isprime}[c] \leftarrow 1$
  \EndFor
  \State // filter the sieve, using each prime in $\var{primes}[]$
  \For{$i \leftarrow 0$ {\bol{to}} $s-1$}
      \label{stp-primes5-filter1}
    \State set $c \leftarrow \var{next}[i]$
    \While{$c < \var{top}$}
      \State // eliminate multiples of $i$\,th prime from sieve
      \State set $\var{isprime}[c - \var{bot}] \leftarrow 0$
      \State set $c \leftarrow c + \var{primes}[i]$
    \EndWhile
    \State // retain next multiple of this $i$\,th prime
    \State set $\var{next}[i] \leftarrow c$
      \label{stp-primes5-filter2}
  \EndFor
  \State // count the values that have not been eliminated
  \For{$c \leftarrow {\var{bot}}$ {\bol{to}} $\min(n-1, \var{top}-1)$}
      \label{stp-primes5-scan1}
    \If{$\var{isprime}[c - \var{bot}] = 1$}
      \State set $\var{nprimes} \leftarrow \var{nprimes}+1$
    \EndIf
  \EndFor
      \label{stp-primes5-scan2}
  \State // prepare for next segment
  \State set $\var{bot} \leftarrow \var{top}$ and
  		$\var{top} \leftarrow \var{top} + \var{segsize}$
    \label{stp-primes5-iterate}
\EndWhile
\State \Return \var{nprimes}
\end{algorithmic}
 \caption{Counting the primes less than $n$, Method~5 (Sieve of
Eratosthenes, partitioned into segments of size $\var{segsize}$).
\label{alg-primes5}}
\end{algorithm}

Algorithm~\ref{alg-primes5} again reduces the resource cost of
computing primes, but this time the resource being reduced is memory
space, not execution time, another important point that your students
need to appreciate as part of their algorithmic understanding.
Algorithm~\ref{alg-primes5} first computes a list of the $n'$ primes
less than or equal to $\sqrt{n}$, and then uses a {\emph{segmented}}
sieve.
For example, an array of $n'=78{,}497$ primes is required to handle
$n=10^{12}$.
A second parallel array stores information about the upcoming
cancellation point for each of those primes, tracking the next
multiple not yet processed, which for a prime $v$ is initially $v^2$
(step~\ref{stp-primes5-setnext}), as noted in connection with
Algorithm~\ref{alg-primes4}.

To prevent overhead costs from dominating, each segment of the sieve
(the constant $\var{segsize}$ in Algorithm~\ref{alg-primes5}) should
contain at least $\sqrt{n}$ elements.
Each sieve segment represents a contiguous block of candidates,
ranging from $\var{bot}$ to $\var{top}-1$, and is filtered for each
of the $n'$ primes.
Once the segment has been filtered (steps~\ref{stp-primes5-filter1}
to~\ref{stp-primes5-filter2}), it is scanned to count the
non-eliminated candidates (steps~\ref{stp-primes5-scan1}
to~\ref{stp-primes5-scan2}).
Finally the segment bounds are stepped upward
(step~\ref{stp-primes5-iterate}), and the next range of values is
processed.
Figure~\ref{fig-transcript-primes5} shows the output from an implementation
(one of the programs accompanying this paper), in this case counting
the primes less than $n=10^{12}$, one trillion.

\begin{figure}[t]
\centering
\begin{minipage}{80mm}
\small
\begin{verbatim}
---------------------------------------------
Primes 5: Segmented sieve, bells and whistles
---------------------------------------------
finding primes up to n    = 1,000,000,000,000 = 1e+12
number of primes < sqrtn  = 78,497
last of those primes      = 999,983
memory space used         = 9.2 MiB
number of primes < n      = 37,607,912,018
density of primes         = 3.76last of those primes      = 999,999,999,989
computation time          = 1767.173 seconds
\end{verbatim}
\end{minipage}
\mycaption{Number of primes less than $10^{12}$ (one trillion)
computed using Method~5 and segments of size $2^{23}$, using less
than {\mb{10}} of memory, and requiring around 30 minutes of
computation on a 2024 MacBook Pro M3.
\label{fig-transcript-primes5}}
\end{figure}

\myparagraph{Measured Execution Times}

\begin{figure}
\centering
\includegraphics[height=\figheight]{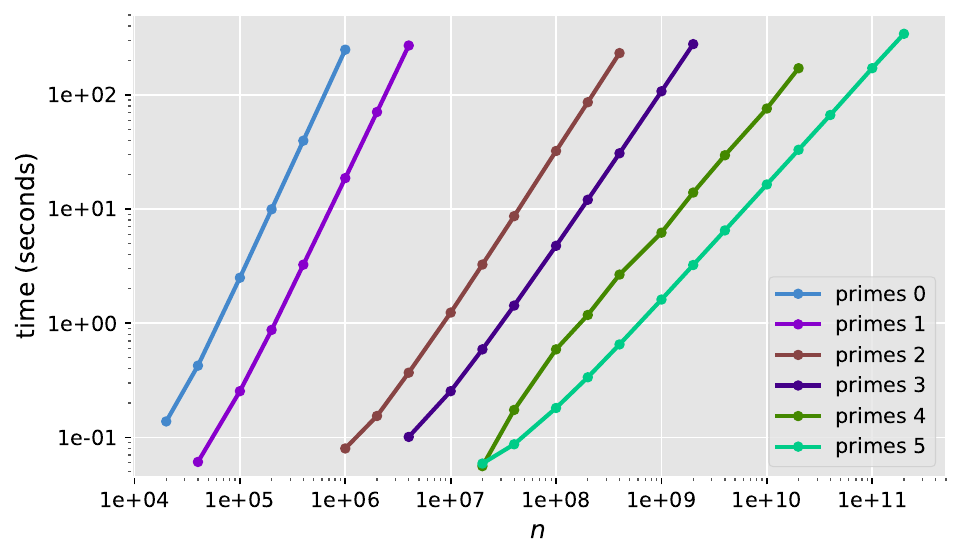}
\mycaption{Execution times for the prime counting task, six different
approaches measured on a 2024 MacBook with an M3 Pro processor and
{\gb{36}} memory.
Method~5 was executed with a segmented sieve of size $2^{23} \approx
8\times 10^6$ one-byte entries.
\label{fig-primes-times}}
\end{figure}

Figure~\ref{fig-primes-times} presents an overview of this
``thrill'' of algorithms, and draws out a number of noteworthy
trends.
These results were collected using the C source code linked to this
paper, executed on a 2024 MacBook Pro with an M3 Pro processor and
{\gb{36}} memory.

Most compelling is the vast discrepancy in capacity between the worst
and best of the methods.
The simplest, Method~0, took $250$ seconds to reach $n=10^6$, whereas
the same amount of time allows Method~5 to exceed $n= 10^{11}$.
That irresistible benefit in terms of available problem size time
arises from the different asymptotic growth rates listed in
Table~\ref{tbl-prime-asymptotics}, and demonstrates that even factors
of $\log n$ provide a worthwhile savings.

Figure~\ref{fig-primes-times}, which uses log-log scales, can also be
used to highlight the fact that different asymptotic growth rates
result in different gradients and not just different starting points.
Extrapolating the lines upward illustrates that the benefit of an
asymptotically superior algorithm magnifies as larger and larger
problems are encountered.

An important third point is illustrated by the speed of Method~5
compared to Method~4.
Method~5 was motivated as saving memory space but using the same
algorithm.
So why does it execute faster?
This question allows you to introduce your students to the notion of
the memory hierarchy.
Because it has a much smaller memory footprint (in Figure, Method~5
was executed using an $\var{isprime}[]$ array of $2^{23}$ bytes and a
total array cost of less than {\mb{10}}) its data structures are
primarily cache resident.
That is, the speed differential between Method~4 and Method~5 is
{\emph{because}} the latter requires much less memory, a rewarding
win-win in terms of computational resources.

\begin{figure}
\centering
\includegraphics[height=\figheight]{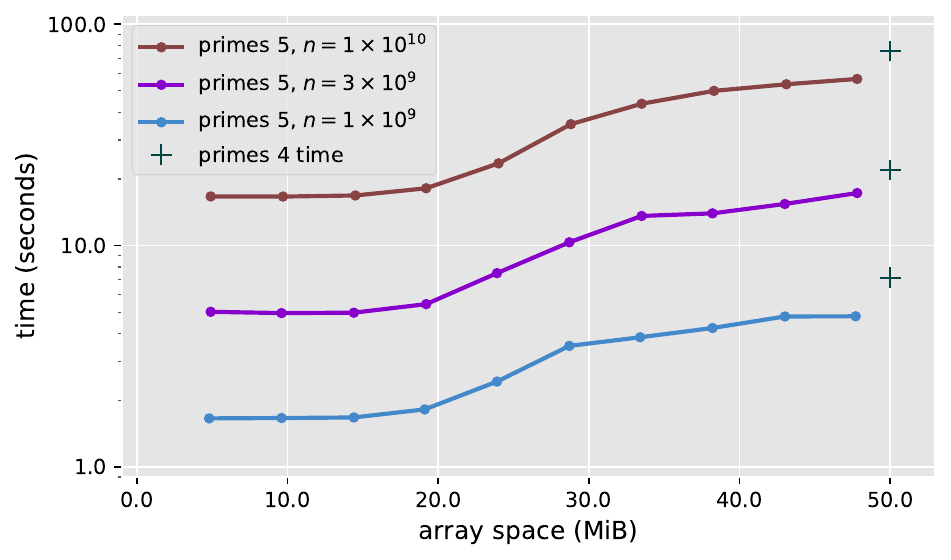}
\mycaption{Execution times for the prime counting task using Method~5
and different segment sizes for three different values of $n$,
plotted as a function of memory space required for arrays, using the
same hardware as Figure~\ref{fig-primes-times}.
The segment sizes varied from $5 \times 10^6$ (left end of curves) to
$5 \times 10^7$ (right end of curves) in steps of $5 \times 10^6$.
The three ``{\sf{+}}'' marks represent the speed (but not memory
cost, which required $n$ bytes in each case) of Method~4.
\label{fig-primes-segmented}}
\end{figure}

Figure~\ref{fig-primes-segmented} further illustrates that point.
The increased execution times through the middle section of the graph
are a direct consequence of the data structures (arrays
$\var{primes}[]$, $\var{isprime}[]$, and $\var{next}[]$)
transitioning from being fully cache resident through to primarily
memory resident.
The L2 cache size on the MacBook Pro (M3 Pro processor) is believed
to be {\mb{16}},\footnote{See
{\url{https://en.wikipedia.org/wiki/Apple_M3}}, accessed 8 August
2024.}
and that belief is supported by Figure~\ref{fig-primes-segmented}.
Note that the three ``{\sf{+}}'' marks for Method~4 are not correctly
placed on the horizontal scale.
The corresponding three memory requirements are (bottom to top)
{\mb{953}}, {\mb{2861}}, and {\mb{9534}}, vastly greater than is
required by Method~5.

\myparagraph{Parallelism}

Method~5 also provides the opportunity to introduce another important
computing concept, that of parallelism.
One of the authors had a slightly older 2023 MacBook Pro laptop with
an M2 Max chip and {\gb{96}} that they were able to dedicate to some
longer experiments, and sought to compute the primes up to one
quadrillion, $n=10^{15}$.
But extrapolating from the results shown in
Figure~\ref{fig-transcript-primes5} (that is, applying the $n\log\log
n$ growth rate) suggested that around $535$ hours would be required
to do that -- more than $22$ days of continuous computation.

Fortunately, Method~5 is readily parallelizable.
With the exception of the handover status embedded in the
$\var{next}[]$ array, there is almost no interaction required between
the sieve segments.
Furthermore, the $\var{next}[]$ array, while convenient in a purely
sequential implementation, is avoidable if mod/div operations are
used to establish the start point for each prime.
With that modification, and provided the each processing thread is
able to deploy its own $\var{isprime}[]$ array, the segments can
operate completely independently, and the multi-threading
capabilities of the Apple M2 and M3 chips exploited.

After some rather tedious debugging to eliminate ``quadrillion-scale
race condition'' problems, that parallel implementation computed the
$29{,}844{,}570{,}422{,}669$ primes less than $n=10^{15}$ using $20$
threads in $355{,}968$ seconds, or $4.12$ days
({\emph{ta daa}}!)

As the candidates increase, the time per candidate increases slowly,
with more primes to be filtered against the segment.
The first trillion candidates were checked in $301$ seconds in
parallel, whereas the one-thousandth trillion (at the end of the
computation) took $361$ seconds.
Given very large amounts of time, or much greater parallelism, it is
likely that the current code would allow counting primes up to the
order of $n=10^{18}$.
Of course, this still falls well short of what has been achieved by
more sophisticated mechanisms\footnote{For example, see the list at
{\url{https://en.wikipedia.org/wiki/Prime_number_theorem}}, accessed
7 August 2024.}, but those mechanisms are also well beyond what
first- and second-year students can be asked to absorb.

\begin{figure}
\centering
\includegraphics[height=\figheight]{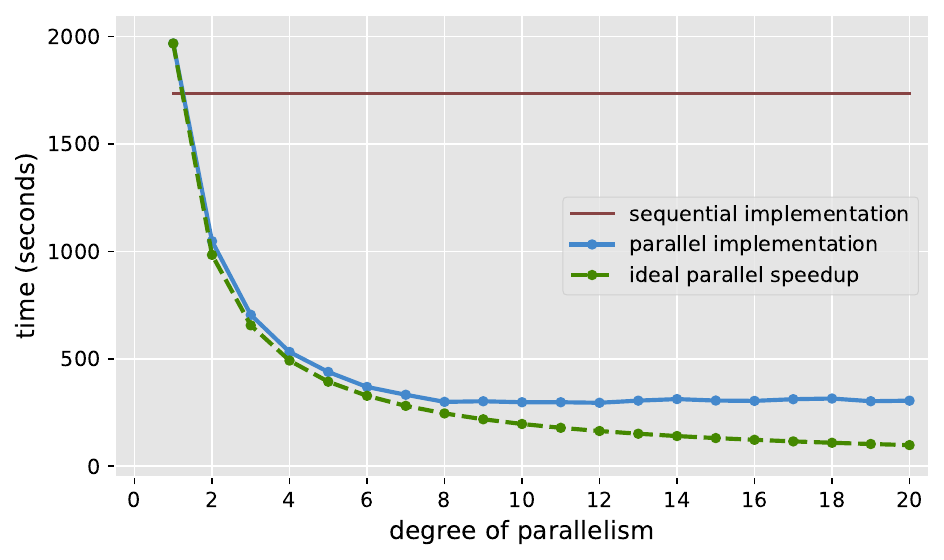}
\mycaption{Execution times for the prime counting task for
$n=10^{12}$ when multiple cores are available and the parallel
version of Algorithm~\ref{alg-primes5} is employed.
\label{fig-parallel-times}}
\end{figure}

Figure~\ref{fig-parallel-times} shows the execution time achieved by
the parallel implementation as the number of processing threads
increases, now using $n=10^{12}$.
Execution times were the average of five observations on the MacBook
Pro M2 Max chip, which has eight ``performance'' cores and four
``efficiency'' ones.
The horizontal brown line in Figure~\ref{fig-parallel-times} shows
the execution time of the sequential version described by
Algorithms~\ref{alg-primes5}; it is faster than the parallel version
with one thread because of its use of the $\var{next}[]$ array, and
because there is an overhead in managing threads.
The dotted green line shows the ideal speedup curve that would result if
perfect parallelism could be attained.
The maximum speed-up achieved was a factor of $6.6$ relative to the
single threaded parallel version.
Attained speed-ups depend upon a range of factors, including the
relationship between the segment size and the size of available cache
memory.
Speed-up was a factor of $5.9$ relative to the pure-sequential
program.
Broadly similar factors were observed in the larger $n=10^{15}$
experiment -- a run that had been estimated as requiring something
like $535$ hours was completed in $136$ hours, for a speedup of
$5.4$.
 \section{Finding Substring Repetitions}
\label{sec-repeats}

We turn now to the question of finding repeated substrings within a
string.
Suppose that $T[0\dots n-1]$ is a string of $n$ characters, where $n$
might be millions or even billions, and that $T[i\dots_m]$ denotes
the $m$-character substring from $T[i]$ to $T[i+m-1]$.
Suppose further that we are concerned that $T[]$ might contain
duplicate substrings, and that we wish to determine the number of
positions $i$ in $T[]$ at which the $m$-substring $T[i\ldots_m]$ is a
repeat of a previous substring $T[j\dots_m]$, for some $j<i$, with
(usually) $m\ll n$.
For example, the string $T={}$``{\tt{merry.mary.marry.me}}'' is of
length $n=19$ and contains multiple repeated substrings of lengths
$m=2$, $m=3$, and $m=4$; and one subsequence of length $m=6$ at
position $i=8$ (the subsequence ``{\tt{ry.mar}}'') which is a repeat
of the corresponding six-symbol substring commencing at position
$j=3$.

\myparagraph{Method 0}

The starting method is again embarrassingly inefficient, and we do
not provide separate pseudocode.
It has a similar relationship to the ``Method~1'' approach that is
described in the next paragraph as was the case between Method~0 and
Method~1 for prime counting, and supposes that its implementor was
unaware of the {\bf{break}} statement, with in this case two being possible
(steps~\ref{stp-repeats1-break1} and~\ref{stp-repeats1-break2} of
Algorithm~\ref{alg-repeats1}).

This first method compares $O(n^2)$ pairs of strings and performs $m$
character comparisons for each such comparison, and it thus requires
$O(n^2m)$ time in all executions.

\myparagraph{Method 1}

\begin{algorithm}[t]
\begin{algorithmic}[1]
\State set $\var{nrepeats} \leftarrow 0$
\State // consider each substring in turn
\For{$i \leftarrow 1$ {\bol{to}} $n-m$}
  \State // and every possible previous substring
  \For{$j \leftarrow 0$ {\bol{to}} $i-1$}
    \State // compare (up to) $m$ characters
    \State set $\var{match} \leftarrow 1$
    \For{$c \leftarrow 0$ {\bol{to}} $m-1$}
      \If{$T[j+c] \not= T[i+c]$}
	\State // not a match, and no need to keep checking
        \State set $\var{match} \leftarrow 0$
	\State {\bol{break}}
	  \label{stp-repeats1-break1}
      \EndIf
    \EndFor
    \If{$\var{match} = 1$}
      \State // $T[i\dots_m]$ is a repeat of $T[j\dots_m]$
      \State set $\var{nrepeats} \leftarrow \var{nrepeats} + 1$
      \State // and no need to keep checking
      \State {\bol{break}}
      \label{stp-repeats1-break2}
    \EndIf
  \EndFor
\EndFor
\State \Return \var{nrepeats}
\end{algorithmic}
 \caption{Method 1 for finding repeated substrings.
\label{alg-repeats1}}
\end{algorithm}

Algorithm~\ref{alg-repeats1} shows a more sensible approach to the
problem of finding repeated substrings.
Each pair of possible paired commencement points $i$ and $j<i$ is
formulated via the outermost and middle loops, and then checked via
the innermost loop.
If the next $m$ characters from $T[j\dots]$ match the next $m$
characters from $T[i\dots]$, then $T[i\dots]$ is a reoccurrence.
The {\bf{break}} statement at step~\ref{stp-repeats1-break1} provides
early exit from the innermost loop as soon as a negative outcome
becomes apparent, an acceleration for the (perhaps relatively common)
case when there are early discrepancies.
That first {\bf{break}} statement means that on random strings the
inner loop rarely executes more than once or twice, and hence that
the expected running time is $O(n^2)$, with $m$ no longer a factor.

The second {\bf{break}}, at step~\ref{stp-repeats1-break1}, triggers
when a match for $T[i\dots_m]$ has been found, and allows the outer
loop to advance to the next string $T[(i+1)\dots_m]$.
This {\bf{break}} saves time on highly repetitive strings with many
substring repetitions.
But worst-case execution time for Method~1 remains $O(n^2m)$, albeit
now for pathologically adversarial input strings $T$ rather than for
all strings.

\myparagraph{Method 2}

\begin{algorithm}[t]
\begin{algorithmic}[1]
\State set $\var{nrepeats} \leftarrow 0$
\State // first, build a suffix array $S[]$ that indexes $T[]$
\For{$i \leftarrow 0$ {\bol{to}} $n-m$}
  \State set $S[i] \leftarrow i$
\EndFor
\State {\bol{sort}} array $S$ so that for all
	$0 \le j < i \le n-m$ 
	\label{stp-repeats2-sort}
\State
	\qquad $T[S[j]\dots_m] \le T[S[i]\dots_m]$
\State // now check consecutive entries in $S[]$, looking for
\State // pairs that have the same $m$-prefix
\For{$i \leftarrow 1$ {\bol{to}} $n-m$}
    \label{stp-repeats2-scanning1}
  \If{$T[S[i-1]\dots_m] = T[S[i]\dots_m]$}
    \State // one or the other of these two is a repeat
    \State set $\var{nrepeats} \leftarrow \var{nrepeats} + 1$
  \EndIf
\EndFor
    \label{stp-repeats2-scanning2}
\State \Return \var{nrepeats}
\end{algorithmic}
 \caption{Method 2 for finding repeated substrings.
\label{alg-repeats2}}
\end{algorithm}

Algorithm~\ref{alg-repeats2} illustrates a quite different approach.
It builds an index data structure known as a {\emph{suffix array}}
{\citep{mm93siamjc}}, a list of indices into $T[]$ that is in sorted
order of the corresponding suffixes in $T$.
The suffix array $S[]$ for a string $T[]$ is built by initializing
$S[i] \leftarrow i$, and then sorting $S[]$ via an indirect string
sort, to achieve an arrangement in which if $j<i$ then
$T[S[j]\ldots]\le T[S[i]\ldots]$.
When building the suffix array it is assumed that the final element
in $T[]$ is a unique occurrence of sentinel character that is the
smallest symbol in the alphabet being used.

Developed as a support tool for pattern search, suffix arrays also
support the repeat finding problem, with a sequential scan of the
sorted suffixes allowing repeated $m$-item prefixes to be identified
(steps~\ref{stp-repeats2-scanning1} to~\ref{stp-repeats2-scanning2}
in Algorithm~\ref{alg-repeats2}).
Moreover, when being applied to the repeat finding task the suffixes
only need to be partially sorted, taking a maximum of $m$ characters
into account, so as to achieve $T[S[j]\ldots_m]\le T[S[i]\ldots_m]$
when $j<i$.
In C programming terms, that means using {\tt{strncmp()}} at
step~\ref{stp-repeats2-sort} in Algorithm~\ref{alg-repeats2} rather
than {\tt{strcmp()}}, with the latter required if a full suffix array
is being built.

The execution time of Algorithm~\ref{alg-repeats2} is dominated by
the sorting step.
If an $O(n\log n)$ worst-case (comparisons) algorithm is employed,
and the input is a random string, then on average sort-adjacent
suffixes diverge after $O(\log n)$ character comparisons, making the
sorting cost $O(n\log^2 n)$ on average, and $O(nm\log n)$ in the
worst case (assuming $m>\log n$).
Moreover, the post-sort scanning process can never cost more than was
spent doing the sorting.

\myparagraph{Method 3}

\begin{algorithm}[t]
\begin{algorithmic}[1]
\Function{\mbox{\rm\var{tqsort}}}{$S, \var{lo}, \var{hi}, \var{depth}$}
\If{$\var{lo} \ge \var{hi}$}
  \State // simplest base case, there are no repeats in this range
  \State {\bol{return}} 0
\EndIf
\If{$\var{depth}=m$}
  \State // second base case, and there may be repeats
    \label{stp-repeats3-width}
  \State {\bol{return}} $\var{hi} - \var{lo}$
\EndIf
\State set $\var{pivot\_char} \leftarrow T[S[p] + \var{depth}]$
	for some $\var{lo}\le p\le\var{hi}$
\State {\bol{reorder}} $S[\var{lo}\ldots\var{hi}]$ using variables \var{eq}
	and \var{fg} so that:
	  \label{stp-repeats3-partition1}
	\\
	\quad\qquad $\var{lo} \le \var{eq} < \var{fg} \le \var{hi}+1$ and
	\\
	\quad\qquad $T[S[p]+\var{depth}] < \var{pivot\_char}$ for
		$\var{lo}\le p < \var{eq}$ and
	\\
	\quad\qquad $T[S[p]+\var{depth}] = \var{pivot\_char}$ for
		$\var{eq}\le p < \var{fg}$ and
	\\
	\quad\qquad $T[S[p]+\var{depth}] > \var{pivot\_char}$ for
		$\var{fg}\le p \le \var{hi}$
	  \label{stp-repeats3-partition2}
\State {\bol{return}}
	\\
	\quad\qquad \var{tqsort}$(S, \var{lo}, \var{eq}-1, \var{depth}) + {}$
	  \label{stp-repeats3-recurse1}
	\\
	\quad\qquad \var{tqsort}$(S, \var{eq}, \var{fg}-1, \var{depth}+1) + {}$
	  \label{stp-repeats3-recurse-eql}
	\\
	\quad\qquad \var{tqsort}$(S, \var{fg}, \var{hi}, \var{depth})$
	  \label{stp-repeats3-recurse2}

\EndFunction

\State
\State // the main computation first initializes a suffix array

\For{$i \leftarrow 0$ {\bol{to}} $n-m$}
  \State set $S[i] \leftarrow i$
\EndFor
\State // then orders the indices within it
\State set $\var{nrepeats} \leftarrow \var{tqsort}(S, 0, n-m, 0)$
\State // and then has no further need of it!
\State \Return \var{nrepeats}
\end{algorithmic}
 \caption{Method 3 for finding repeated substrings, using ternary
quicksort {\citep{bs97soda}} to order the strings.
The repeat counting is directly embedded in the sorting process.
\label{alg-repeats3}}
\end{algorithm}

The final repeat-finding approach we employ in our lectures also
computes an approximate-to-$m$ suffix array, but obtains the repeat
count as a direct product of the suffix sorting process, see
Algorithm~\ref{alg-repeats3}.
This mechanism is based on the {\emph{ternary quicksort}} of
{\citet{bs97soda}}, a combination of radix sorting and quicksort.
Rather than sorting by comparing whole strings -- which as the
sorting process converges will tend to have increasingly longer
identical prefixes -- this approach considers only characters at
position $\var{depth}$ during each partitioning/recursion cycle,
where $\var{depth}$ is a parameter to the recursive call.
The pre-condition associated with $\var{depth}$ that all of the
suffixes $S[\var{lo}\ldots\var{hi}]$ in the current range are
identical at positions $0$ to $\var{depth}-1$.

Provided that $\var{depth}$ is still less than $m$ (the base case at
step~\ref{stp-repeats3-width}), this current call then performs a
three-way partition using the character at position $\var{depth}$ in
each of the strings (steps~\ref{stp-repeats3-partition1}
to~\ref{stp-repeats3-partition2} in Algorithm~\ref{alg-repeats3}), to
form three zones, one where all the strings are less than the pivot
character at position $\var{depth}$, one where all the strings are
equal to the pivot character at position $\var{depth}$, and one where
all of the strings are greater.
The first and last groups might be empty, but the middle group must
contain at least one suffix.

Three recursive calls are then made
(steps~\ref{stp-repeats3-recurse1} to~\ref{stp-repeats3-recurse2}),
with the equal zone deepening to position $\var{depth}+1$
(step~\ref{stp-repeats3-recurse-eql}), in keeping with the invariant.
Not only is the sorting process faster, but also the post-sorting
linear scan of Method~2 (and the $n-1$ {\tt{strncmp()}} calls that
are required by it) is avoided, since the hierarchy of recursive
calls has already carried out the required counting.
That is, the initial call $\var{tqsort}(S, 0, n-m, 0)$ both
constructs the depth-$m$ approximate suffix array $S$ and
simultaneously computes the required answer, with the suffix array
not actually used once it has been constructed.

Like quicksort itself, in the worst case this approach can require
$O(n^2)$ time (or, more precisely, $O(n\cdot\min(n, m\sigma))$, where
$\sigma$ is the size of the alphabet over which the strings have been
formed), a cost that arises if a carefully constructed adversarial
input is targeted to the pivot-selection and partitioning processes.
But if the pivot characters are selected using random sampling at
each recursive call, and/or the input string is typical rather than
adversarial, this approach requires $O(n(m+\log n))$ character
comparisons and thus $O(n(m+\log n))$ time {\citep{bs97soda}}.

While the analysis is more complex than might be appropriate for
first-year algorithms students, Method~3 itself is well within their
grasp, and provides a compelling example of recursive
thinking.

\myparagraph{Measured Execution Times}

\begin{figure}[t]
\centering
\includegraphics[height=\figheight]{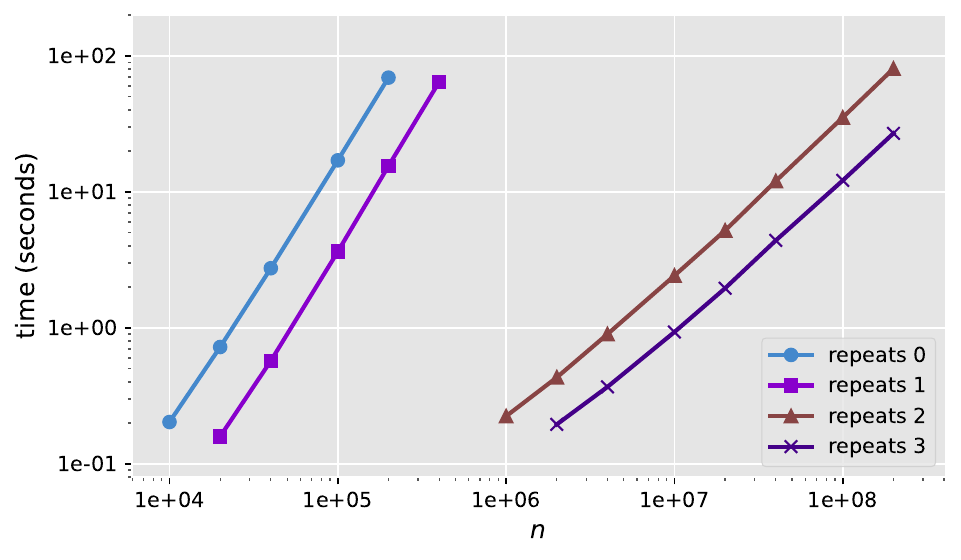}
\mycaption{Execution times for the repeated substring counting task,
four different approaches measured on a 2024 Macbook with an M3 Pro
processor and {\gb{36}} memory.
The match length $m=50$ was used throughout, and only $n$ was varied.
\label{fig-repeats-times}}
\end{figure}

Figure~\ref{fig-repeats-times} plots execution times for the
substring repetition task, drawing out the dramatic differences in
execution times between the four methods.
All execution times are for $m=50$ as $n$ increases, with sequences
of $n$ characters taken as a prefix of a large text file containing
$266.6$~MiB of newspaper text with embedded SGML markup tags, sourced
under license from the Wall Street Journal and covering 1987, 1988,
and 1989.\footnote{See ``Text Research Collection Volume 1, Revised
March 1994'' at {\url{https://trec.nist.gov/data/docs_eng.html}} for
the provenance of this data file, most recently accessed 29 September
2024.}

\begin{figure}[t]
\centering
\begin{minipage}{80mm}
\small
\begin{verbatim}
-----------------------------------------------------
Repeats 3: Build suffix array using ternary quicksort
-----------------------------------------------------
text size n         = 200,000,000 = 2e+08 from wsj.txt
repeat length m     = 50
memory space used   = 1525.9 MiB
number of m-repeats = 10,189,540
computation time    = 26.253 seconds
\end{verbatim}
\end{minipage}
\mycaption{Finding repeated substrings in $200$ million bytes of
newspaper text using Repeats Method~3 (Algorithm~\ref{alg-repeats3}),
executed on a 2024 MacBook Pro M3.
More than $5$\% of the $50$-character substrings are repeats.
\label{fig-transcript-repeats3}}
\end{figure}

Figure~\ref{fig-transcript-repeats3} shows the output of an
implementation of Method~3, corresponding to the rightmost purple
``{\sf{x}}'' in Figure~\ref{fig-repeats-times}.
Repeated strings of length $m=50$ are relatively common, in part a
consequence of the document labeling and markup scheme used in this
text.
When $m$ is increased to $250$ the number of repeated substrings
drops to $13{,}766$.
Note also that there is a non-trivial space burden associated with
the suffix array, and in this implementation it requires an eight
byte pointer per byte of underlying string.
This is another common algorithmic theme -- the benefit of exchanging
increased space for decreased computation time.

Figure~\ref{fig-tadaa} is from a lecture capture, showing Method~2
executing an $n=10{,}000{,}000$ and $m=100$ computation in just $3$
seconds, compared to an extrapolation of Method~1 requiring nine
hours for the same task.
Method~1 can be similarly extrapolated as requiring approximately six
months for the computation shown in
Figure~\ref{fig-transcript-repeats3}.
The difference between the two is visceral.

 \section{Related Work}

There has been a variety of literature in which methods for conveying
algorithmic knowledge are considered.
Here we comment on just a fraction of what is available.

In past work {\citet{sigcse01gt}} suggest that algorithmics education
can be motivated to students by internet examples, including graph
algorithms and inverted index structures; and
{\citet{sigcse02saunders}} argues for empirical analysis as an
important adjunct, taking a position that is in similar to our use of
measured execution times, albeit at an honors level.

{\citet{csej04gvz}} describe an approach whereby they present
students with functions of $O(n^2)$ complexity involving integer
arrays, and then assign follow-up tasks such as determining the
complexity, and then giving an improved function.
In the exam, the students are given a further function and expected
to perform the same tasks.
Student exam performance increased as a result of the suggested
methodology.

More recently, {\citet{sigcse16fhlmw}} argue that an
``active'' approach to learning algorithms is helpful, describing an
experiment and measurements arising from their use of an active eBook
that includes animations.
By running code live in lectures, and waiting patiently (or
impatiently, as the case may be) for it to finish, and building a
spreadsheet that tabulates and predicts, we are seeking to have
``brain active'' lectures that help reach those same goals.

{\citet{gph16}} suggest the use of virtual reality and gamification
to make learning of search algorithms more entertaining, leading to
higher student motivation and learning efficiency.
{\citet{sigcse21szdg}} report poor student motivation in studying DSA
and, like {\citet{gph16}} use 2.5D visualizations and game-based
learning to improve motivation and learning outcomes.
{\citet{tmcs16vs}} also report the use of animations and didactic
games to teach algorithms and algorithmic efficiency and report
improvements.
In a related vein, {\citet{gjch21}} argue that traditional complexity
analysis has limited value in motivating students about the
importance of algorithmic efficiency.
They propose instead that student concern with respect to climate
change be harnessed, and that suggest a real world comparison of
energy consumption by different algorithms.
{\citet{ace21az}} synthesize the outcomes of interviews with five
experts and fourteen novices, seeking to understand how they approach
sections of code when challenged to describe their algorithmic
complexity, observing that novices were less likely to note
interactions between nested loops, and more likely to regard them as
being orthogonal and hence multiplying their individual execution
times.

{\citet{sigcse24fhlmw}} explore what students and teachers understand
the word ``algorithm'' to mean.
This accords with another element of our approach, that emphasizes
that each problem has many applicable algorithms, and that they may
have different attributes of which running time is one, but space
used and ease of implementation are others.
Concurrently, {\citet{sigcse24lpcwgf}} analyze data collected in a
literature survey, and present an overall taxonomy of 97 identified
papers, including which algorithmic topics were explored in that
work, what experimental structure was used, and what was measured.
Interestingly, neither of the two tasks we consider here appear in
their breakdown; perhaps being too foundational.
In our case we start with the foundational, so as to be able to move
relatively quickly to the hierarchy of algorithms.
In a different line of work, {\citet{ace24vanrenssen}} argues that
students benefit from examples and questions of gradated difficulty,
reflecting on experience in a subject that sounds much like the one
we consider here.
That aligns with our experience, and reinforces the usefulness of
starting each ``thrill'' with an obvious approach.

While much of this previous work has considered how to make the
teaching and learning of algorithms effective, none has considered
how to make it {\emph{exciting}}, which is the particular emphasis we
have sought to present in this work.
Physics students know well the astonishing impact that a practical
demonstration of a (for example) Van de Graaff generator can have in a
lecture, and the noise that can be generated by a 10~kV or higher
static discharge.
That is the electric (forgive the pun) atmosphere that we also seek
to bring to our early lectures on algorithms.

 \section{Conclusion}
\label{sec-conclusion}

We have described a way of conveying the visceral thrill of
algorithmic improvements, and provided two case studies that are
suitable for presentation to students in a ``foundations of
algorithms'' subject, once they have mastered the fundamentals of
programming in their first subject.
There is nothing novel in any of the methods we present, and all of
them are well-known to the point where there are multiple versions of
them on the public web (which means that they should not be assigned
as programming assignments, of course).
There are also further algorithms for solving these two tasks -- the
ones shown here are by no means the last word in the story (see, for
example, {\citet{joa83pritchard}} for prime counting, and
{\citet{ieeetc11nzc}} for suffix sorting).

Nevertheless, we argue that chaining a sequence of solutions into a
hierarchy as we have done here, starting with a ``dumb brute force''
approach can provide an ``excitement'' that engages undergraduate
computing students, and sets them up for their subsequent study.
Almost every other computation task has an ``obvious'' way and then a
step-ladder of ``better'' ways too, including sorting and a wide
range of graph and search problem.
The particular attraction of the two that we have described here is
that the hierarchy of solutions -- the ``thrill'' -- is also easy to
grasp based on foundational programming skills.

\myparagraph{Software}
You'll learn a lot by writing your own versions of the software, much
more than if you simply run our code.
But if you are desperately short of time, try emailing the first
author.

\balance \renewcommand{\bibsep}{3.5pt}
\bibliographystyle{abbrvnat}

\end{document}